\documentclass[12pt]{article}
\usepackage[utf8]{inputenc}
\usepackage{amsmath,latexsym,amssymb,hyperref,graphicx}
\usepackage{authblk}

\begin{document}

\title{A $SU(5)\times Z_2$ kink solution\\  and its local stability}
\author[$\dagger$]{Rommel Guerrero}
\author[$\dagger$]{R. Omar Rodriguez}
\author[$\star$]{Rafael Chavez}
\affil[$\dagger$]{Facultad de Ciencias, Escuela Superior Polit\'ecnica de Chimborazo, EC060155-Riobamba, Ecuador}
\affil[$\star$]{Departamento de Ciencias B\'asicas, Universidad Polit\'ecnica Salesiana, 170105-Quito, Ecuador}

\maketitle

\begin{abstract}
A non-abelian kink inducing asymptotically the breaking pattern $SU(5)\times Z_2\rightarrow SU(4)\times U(1)/Z_4$ is obtained. We consider a fourth order Higgs potential in a $1+1$ theory where the scalar field is in the adjoint representation of $SU(5)$. The perturbative stability of the kink  is also evaluated. A Schr\"odinger-like equation for the excitations along each $SU(5)$ generator is determined, and in none of the cases negative eigenvalues compromising the stability of solution are found. In particular, several bounded scalar states are found, being one of them the translational zero mode of the flat space $SU(5)\times Z_2$ kink.\\


\end{abstract}

\section{Introduction}

In theories with a simple scalar field and a $Z_2$ invariant self-interaction potential, it is possible to find stable topological kinks interpolating asymptotically between the minima of the system.  These solutions are interesting  in the framework of the gravitational theories with extra dimensions because the kink or domain wall interpolates between two Anti de Sitter spacetimes and induces the standard gravitational interaction on the four-dimensional sector of the warped structure \cite{Randall:1999vf,Gremm:1999pj,Bajc:1999mh,CastilloFelisola:2004eg}. Such scenarios are referred to as brane-worlds, and it is surmised that  the standard model fields should be localized on the topological defect \cite{Melfo:2006hh,Guerrero:2006gj,Guerrero:2009ac}. 
 
As a first approximation to a scenario where the symmetry group of the standard model can be recovered inside the domain wall, in absence of gravity but in presence of a non-abelian symmetry, kink solutions  have been obtained in several opportunities \cite{Shin:2003xy,Davidson:2007cf,Chavez:2016sbv}. In Ref. \cite{Shin:2003xy} three kink solutions for a $SO(10)$ theory  inducing asymptotically the breaking symmetry $SO(10)\rightarrow SU(5)$ were determined; subsequently, in Ref. \cite{Chavez:2016sbv} the local stability of these scenarios was evaluated finding in two of them tachyonic P\"oschl-Teller modes in the spectrum of scalar perturbations. Other models in terms of $E_6$ group were discussed in Ref. \cite{Davidson:2007cf}.





Among the non-abelian solutions we highlight the one where the kink interpolates between $SU(3)\times SU(2)\times U(1)$ vacuum expectation values (vev) of a $SU(5)\times Z_2$ invariant potential \cite{Pogosian:2000xv,Vachaspati:2001pw,Melfo:2011ev,Pantoja:2015yin}. Since this issue is the focus of this paper, let us review in detail this  scenario.


Consider the bosonic sector of the $SU(5)$ model in $(1+1)$ dimensions
\begin{equation}\label{lagrangian}
L=-\text{Tr}(\partial_m{\bf\Phi}\partial^m{\bf\Phi})-V({\bf\Phi}),
\end{equation}
\begin{equation}\label{potential}
V({\bf\Phi})=-\mu^2\text{Tr}({\bf\Phi}^2)+h(\text{Tr}({\bf\Phi}^2))^2+\lambda\text{Tr}({\bf\Phi}^4)+V_0
\end{equation} 
with ${\bf \Phi}$ a scalar field transforming in the adjoint representation of the symmetry group 
\begin{equation}
{\bf\Phi}\rightarrow{\bf U}{\bf \Phi}{\bf U}^{\dagger},\quad {\bf U}=\exp(i\alpha_j{\bf T}_j),\quad \text{Tr}({\bf T}_{j_1} {\bf T}_{j_2 })=\frac{1}{2}\delta_{j_1j_2}
\end{equation}
where ${\bf T}_j$, $j=1,\dots, 24$, are  traceless hermitian generators of $SU(5)$. In the potential (\ref{potential}), $\mu$, $h$ and $\lambda$ are the parameters of the theory and $V_0$ is a constant to adjust conveniently the minimum to zero. 

It is well know that there are two possible non-trivial form for the minimum of  (\ref{potential}) \cite{Li:1973mq}: 
\begin{equation}\label{vA}
<{\bf\Phi}_{\text A}>\sim\text{diag}(2,2,2,-3,-3),\qquad \lambda>0,
\end{equation}
and 
\begin{equation}\label{vevB}
<{\bf\Phi}_{\text B}>\sim\text{diag}(1,1,1,1,-4),\qquad \lambda<0;
\end{equation}
which lead to the breaking patterns  
\begin{equation}
SU(5)\rightarrow SU(3)\times SU(2)\times U(1)
\end{equation}
and
\begin{equation}\label{vevB1}
SU(5)\rightarrow SU(4)\times  U(1), 
\end{equation}
respectively.

Due to the absence of cubic terms in (\ref{potential}) the $Z_2$ symmetry is included in the model and kink solutions are expected, such that
\begin{equation}\label{+inf-}
\boldsymbol{\Phi}(z=-\infty)=-{\bf U}\boldsymbol{\Phi}(z=+\infty){\bf U}^{\dagger},
\end{equation}
where $U$ is an element of $SU(5)$ and ${\bf\Phi}$ depends only on
the coordinate $z$. In fact,  for $h=-3\lambda/20$ and $\lambda>0$, the symmetry breaking pattern 
\begin{equation}
SU(5)\times Z_2\rightarrow \frac{SU(3)\times SU(2)\times U(1)}{Z_3\times Z_2}
\end{equation}
can be induced asymptotically by the following non-abelian kink
\begin{eqnarray}
{\bf \Phi}_{\text A}&=&\frac{\sqrt{5}}{2}\frac{\mu}{\sqrt{\lambda}}\Big[\text{diag}(1,-1,0,1,-1)\nonumber\\&+&\frac{1}{5}\tanh\Big(\frac{\mu z}{\sqrt{2}}\Big)\text{diag}(-1,-1,4,-1,-1)\Big].\label{campoA}
\end{eqnarray}
which, at $z=\pm\infty$ goes to
\begin{equation}\label{bA+}
{\bf \Phi}_{\text A}(z=+\infty)=\frac{1}{\sqrt{5}}\frac{\mu}{\sqrt{\lambda}}\text{diag}(2,-3,2,2,-3)
\end{equation}
\begin{equation}\label{bA-}
{\bf \Phi}_{\text A}(z=-\infty)=\frac{1}{\sqrt{5}}\frac{\mu}{\sqrt{\lambda}}\text{diag}(3,-2,-2,3,-2).
\end{equation}
Notice that (\ref{bA-}) is compatible with the constraint (\ref{+inf-}) and that, from trace properties, the vacuums (\ref{vA}) and (\ref{bA+}) are equivalent for the scalar potential. On the other hand, in accordance with (\ref{bA+}) and (\ref{bA-}), $SU(3)\times SU(2)\times U(1)$ is embedded asymptotically in $SU(5)$ in different ways. Moreover, inside the kink, ${\bf\Phi}_{\text A}(z=0)\sim\text{diag}(1,-1,0,1,-1)$, the unbroken group is
\begin{equation}
\frac{SU(2)\times SU(2)\times U(1)\times U(1)}{Z_2\times Z_2}.
\end{equation}

This solution, as well as its perturbative stability, were determined in \cite{Pogosian:2000xv}; subsequently, (\ref{campoA}) was recovered as a particular case of a kink in $SU(N)\times Z_2$ \cite{Vachaspati:2001pw}, the extension to curved spacetime in five dimensions was  found in \cite{Melfo:2011ev,Pantoja:2015yin} where (\ref{campoA}) induces the symmetry group of the standard model at the boundary of an  AdS${}_5$ warped spacetime.  

With respect to the vev (\ref{vevB}), curiously, up to now, a kink solution for the model (\ref{lagrangian}) has not been reported; however, a self-gravitating kink inducing unbroken group (\ref{vevB1}) was obtained in \cite{Melfo:2011ev}.  Considering the absence of a flat kink for (\ref{vevB}), our proposal for this paper is to find this solution and evaluate its stability under small perturbations.

\section{Kink solution}
Let us consider a kink solution, ${\bf\Phi}_{\text B}$, for (\ref{lagrangian}) in correspondence, at infinity,  with the symmetry breaking pattern 
\begin{equation}\label{SU4U1}
SU(5)\times Z_2\rightarrow \frac{SU(4)\times U(1)}{Z_4}.
\end{equation}

For this case, it is convenient to write the non-abelian field in terms of  diagonal generators of $SU(5)$
\begin{equation}\label{scalarfield}
{\bf\Phi}= \phi_1{\bf T}_{21}+\phi_2{\bf T}_{22}+\phi_3{\bf T}_{23}+\phi_4{\bf T}_{24}
\end{equation}
where
\begin{eqnarray}
{\bf T}_{21}&=&\frac{1}{2}\text{diag}(1,-1,0,0,0),\\
{\bf T}_{22}&=&\frac{1}{2\sqrt{3}}\text{diag}(1,1,-2,0,0),\\
{\bf T}_{23}&=&\frac{1}{2}\text{diag}(0,0,0,1,-1),\\
{\bf T}_{24}&=&\frac{1}{2\sqrt{15}}\text{diag}(2,2,2,-3,-3).
\end{eqnarray}

The unbroken group (\ref{SU4U1}) is induced by a kink  when (\ref{scalarfield}) satisfy the boundary conditions 
\begin{equation}
{\bf\Phi}(z=+\infty)=v\Big( {\bf T}_{23}+\sqrt{\frac{3}{5}}{\bf T}_{24}\Big)=<{\bf\Phi}_{\text B}>,\label{boundary+}
\end{equation}
\begin{equation}
{\bf \Phi}(z=-\infty)=v\Big( {\bf T}_{23}-\sqrt{\frac{3}{5}}{\bf T}_{24}\Big)=-{\bf U}<{\bf\Phi}_{\text B}>{\bf U}^{\dagger}.\label{boundary-}
\end{equation}

The equations of motion for the coefficients of (\ref{scalarfield}) are given by \begin{equation}
\phi_1^{\prime\prime}=-[\mu^2-(h+\frac{2\lambda}{5})\phi_4^2-(h+\frac{\lambda}{2})(\phi_1^2+\phi_2^2)-h\phi_3^2]\phi_1+\frac{2\lambda}{\sqrt{5}}\phi_1\phi_2\phi_4,
\end{equation}
\begin{equation}
\phi_2^{\prime\prime}=-[\mu^2-(h+\frac{2\lambda}{5})\phi_4^2-(h+\frac{\lambda}{2})(\phi_1^2+\phi_2^2)-h\phi_3^2]\phi_2 +\frac{\lambda}{\sqrt{5}}\phi_4(\phi_1^2-\phi_2^2),
\end{equation}
\begin{equation}
\phi_3^{\prime\prime}=-[\mu^2-(h+\frac{9\lambda}{10})\phi_4^2-(h+\frac{\lambda}{2})\phi_3^2-h(\phi_1^2+\phi_2^2)]\phi_3,
\end{equation}
\begin{equation}
\phi_4^{\prime\prime}=-[\mu^2-(h+\frac{7\lambda}{30})\phi_4^2-(h+\frac{2\lambda}{5})(\phi_1^2+\phi_2^2)-(h+\frac{9\lambda}{10})\phi_3^2]\phi_4+\frac{\lambda}{\sqrt{5}}\phi_2(\phi_1^2-\frac{\phi_2^2}{3}).
\end{equation}
where prime means derivative with respect to $z$. Suggested by (\ref{boundary+}, \ref{boundary-}) we choose conveniently $\phi_1=\phi_2=0$; thus, we obtain a pair of coupled equations for $\phi_3$ and $\phi_4$, namely
\begin{eqnarray}
\phi_3^{\prime\prime}&=&[-\mu^2+(h+\frac{9\lambda}{10})\phi_4^2+(h+\frac{\lambda}{2})\phi_3^2]\phi_3,\label{eqc}\\
\phi_4^{\prime\prime}&=&[-\mu^2+(h+\frac{7\lambda}{30})\phi_4^2+(h+\frac{9\lambda}{10})\phi_3^2]\phi_4\label{eqd},
\end{eqnarray}
which  can be decouple
for $\lambda=-10 h/9$, $h>0$.  Now, for the remaining equations we require 
\begin{equation}
\phi_3(z=\pm\infty)=v,\qquad
\phi_4(z=\pm\infty)=\pm\sqrt{\frac{3}{5}}v;
\end{equation} 
thus, we find the solution
\begin{equation}\label{kinksolution}
{\bf\Phi}_{\text B}=v\left[ {\bf T}_{23}+\sqrt{\frac{3}{5}}\tanh\left(\frac{\mu z}{\sqrt{2}}\right){\bf T}_{24}\right], \quad v=\frac{3\mu}{2\sqrt{h}}
\end{equation}
such that at $z=\pm\infty$,  the unbroken group (\ref{SU4U1}) is embedded  in $SU(5)$ in different ways. On the other hand, in the core of the kink, $z=0$, the remaining symmetry group is
\begin{equation}\label{insidekink}
\frac{SU(3)\times U(1)\times U(1)}{Z_3}.
\end{equation}

\section{Perturbative stability}

Now, in order to study the perturbative stability of  (\ref{kinksolution}) let us consider, in the energy of the system,  small deviations from kink solution
\begin{equation}\label{E}
E=\int dz\left[\text{Tr}(\partial_z{\bf\Phi}_{\text B}+\epsilon\ \partial_z{\bf \Psi})^{2}+V({\bf\Phi}_{\text B}+\epsilon{\bf \Psi})\right],\quad \epsilon\ll 1,
\end{equation}
where $\mathbf{\Psi}=\psi_j {\bf T}_j$. Thus,  to second order in $\epsilon$ (the term proportional to $\epsilon$ is zero via the equation of motion of ${\bf \Phi}_{\text B}$), we find that
\begin{equation}\label{Ep}
E=E[{\bf \Phi}_{\text B}]+\epsilon^2\int dz\ \psi_{j_1}\left(-\delta_{j_1 j_2}\partial_z^2+V_{j_1 j_2}\right)\psi_{j_2}+{\cal O}(\epsilon^3)
\end{equation}
where
\begin{eqnarray}
V_{j_1 j_2}({\bf\Phi}_{\text B})=&-&\frac{1}{2}\mu^2\delta_{j_1 j_2}+h\text{Tr}({\bf\Phi}_{\text B}^2)\delta_{j_1 j_2}+4h\text{Tr}({\bf\Phi}_{\text B}{\bf T}_{j_1})\text{Tr}({\bf\Phi}_{\text B}{\bf T}_{j_2})\nonumber\\&-&\frac{40}{9}h\text{Tr}({\bf\Phi}_{\text B}^2{\bf T}_{j_1}{\bf T}_{j_2})-\frac{20}{9}h\text{Tr}({\bf\Phi}_{\text B}{\bf T}_{j_1}{\bf\Phi}_{\text B}{\bf T}_{j_2})
\end{eqnarray}
which is diagonal and, hence, the double sum in (\ref{Ep}) can be written as follows
\begin{equation}\label{sum}
\left(-\frac{1}{2}\delta_{j_1j_2}\partial_z^2+V_{j_1j_2}\right)\psi_{j_2} =m_{j_1j_2}^2\psi_{j_2}^2,
\end{equation}
and the stability problem consists in determining the eigenvalue associated with each generator of $SU(5)$. Fortunately,  for several generators we obtain the same eigenvalues equation and, after group them, only  five non trivial cases need to be checked. For the trivial cases, $j=19, 20$, whose generators  are broken everywhere, a vanishing potential is obtained. 

For the remaining five cases it
is convenient to consider $\xi=\mu z/\sqrt{2}$, to make dimensionless the equations of motions. For the set of generators labelled by $j=1,\dots, 6$, $ 21, 22$, basis for $SU(3)$ in (\ref{insidekink}),  we get
\begin{equation}\label{ecs1}
\left[-\frac{1}{2}\partial_\xi^2+\frac{1}{4}\left(5+3\tanh^2\xi\right)\right]\psi_j=2\frac{m_j^2}{\mu^2}\psi_j , 
\end{equation}
which can be rewritten in terms of a  P\"oschl-Teller potential,  whose bound states are well known \cite{Poschl:1933zz}. Thus, the spectrum of scalar states is determined for all $j$ by a pair of normalizable eigenfunctions
\begin{equation}
m_0^2=\frac{4+\sqrt{7}}{8}\mu^2\ ,\quad \psi_0\sim \cosh^{(1-\sqrt{7})/2}(\xi)
\end{equation}
\begin{equation}
m_1^2=\frac{3\sqrt{7}}{8}\mu^2\ ,\quad \psi_1\sim \cosh^{(1-\sqrt{7})/2}(\xi)\sinh(\xi)
\end{equation}
and a set of free states from $m^2>3\sqrt{7}\mu^2/8$.

\begin{figure}
\begin{center}
\includegraphics[width=8.5cm,angle=0]{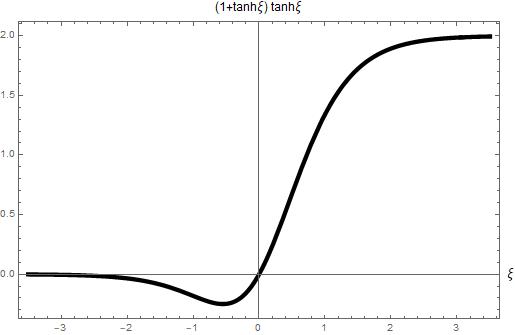}
 \caption{ Plot of the potential for the scalar perturbations $\psi_{_{j_+}}$ along the broken generators $j_+=7,\dots, 12$. The potential profile for $\psi_{_{j_-}}$ with $j_-=13,\dots,18$ is a mirror image of the shown one.}
\label{fig1}
\end{center}
\end{figure}

In the cases identified  with $j_+=7,\dots, 12$ and $j_-=13,\dots,18$, broken generators with respect to (\ref{insidekink}) but unbroken with respect to
(\ref{SU4U1}) at $\xi\rightarrow \pm\infty$, we have 
\begin{equation}\label{eq2case}
\left[-\frac{1}{2}\ \partial_\xi^2+\tanh\xi\left(\tanh\xi \pm 1\right)\right]\psi_{j_\pm}=2\frac{m_{j_\pm}^2}{\mu^2}\psi_{j_\pm}.
\end{equation}
In this case the Schr\"{o}dinger-like equation is associated to a non-conventional potential (see Fig.\ref{fig1}) which  exhibits a negative well in $-\infty<\xi\leq 0$ for $\psi_+$. For $\psi_-$ the potential has an equivalent profile, but in the region $\infty>\xi\geq 0$. However, the eigenvalues of the equation are positive  since (\ref{eq2case}) can be factorized as follows \cite{Sukumar:1986bq}
\begin{equation}
\left(-\partial_\xi+\beta_{\pm}\right)\left(\partial_\xi+\beta_{\pm}\right)\psi_{j_\pm}=4\frac{m_{j_\pm}^2}{\mu^2}\psi_{j_\pm},
\end{equation}
where
\begin{equation}
\beta_\pm=\mp2\frac{1+3\cosh2\xi\mp\sinh2\xi+2(\tanh\xi\mp1)\xi}{4+e^{\pm2\xi}+e^{\mp2\xi}(3\pm4\xi)}.\end{equation}
In addition, the zero mode can be obtained, $\psi_0\sim 1\mp\tanh\xi$, which is out of the spectrum of eigenfunctions because it is not normalizable, as expected in concordance with Fig. \ref{fig1}.

Finally, in the scalar field directions ${\bf T}_{23}$ and ${\bf T}_{24}$, we find 
\begin{equation}\label{eq23}
\left(-\frac{1}{2}\partial_\xi^2+2\right)\psi_{23}=2\frac{m^2_{23}}{\mu^2}\psi_{23}
\end{equation}
\begin{equation}\label{eq24}
\left(-\frac{1}{2}\partial_\xi^2+3\tanh^2\xi -1\right)\psi_{24}=2\frac{m^2_{24}}{\mu^2}\psi_{24},
\end{equation}
with eigenvalues bounded as $m_{23}^2>\mu^2$ and $m_{24}^2\geq 0$. In the last case, two localized states are found
\begin{equation}\label{zeroT}
m_0^2=0\ ,\quad \psi_0\sim \cosh^{-2}(\xi)
\end{equation}
\begin{equation}
m_1^2=\frac{3}{4}\mu^2\ ,\quad \psi_1\sim \cosh^{-2}(\xi)\sinh(\xi) ,
\end{equation}
being the first one, (\ref{zeroT}), the translational mode of the flat space $SU(5)\times Z_2$ kink \cite{Pogosian:2000xv}. 

Since in all cases the eigenvalues are positive,  the perturbations do not induce instability on  (\ref{kinksolution})  and hence the non-abelian flat domain wall $\mathbf{\Phi}_{\text B}$ is a locally stable solution of (\ref{lagrangian}, \ref{potential}).

\section{Ending comments}

We have derived a flat $SU(5)\times Z_2$ kink interpolating
asymptotically between Minkowskian vacuums with the symmetry breaking pattern $SU(5)\times Z_2\rightarrow SU(4)\times U(1)/Z_4$ and with unbroken group $SU(3)\times U(1)\times U(1)/Z_3$
in the core of the wall.  

With regard to the spectrum of scalar fluctuations, we do not find tachyonic modes compromising the local stability of the non-abelian wall. In particular, we find the translational zero mode and several P\"oschl-Teller confined scalar states along $SU(3)$ basis.

Non-abelian kinks as brane worlds is the next issue that we would like to study. The main problem is that we need to find a scenario where the symmetry group of the standard model corresponds to the unbroken symmetry inside the kink. In our opinion, solutions similar to  (\ref{campoA}) and (\ref{kinksolution}) are a first approximation to this open problem.  


As commented in the introduction, another option has been already explored in \cite{Shin:2003xy,Davidson:2007cf} where the symmetry of the theory is determined by the $SO(10)$ group. In this case, the unbroken symmetry for finite $z$ is achieved used the clash-of-symmetries mechanism. Thus, $SU(3)\times SU(2)\times U(1) \times U(1)$ may be obtained
in the core of a $SO(10)$ wall, which is almost the symmetry expected for a more realistic model.

\section{Acknowledgements}
We  wish to thank Adriana Araujo for her collaboration to complete
this paper. This work was partially financed by UPS project. R. Guerrero and R. O. Rodriguez wish to thank ESPOCH for hospitality during the completion of this work.

\end{document}